\documentstyle[prb,aps,multicol]{revtex}
\begin{document}
\title{First-principles prediction of structure, energetics,
formation enthalpy, elastic constants, polarization, and piezoelectric
constants of AlN, GaN, and InN: comparison of local and
gradient-corrected density-functional theory} 
\author{Agostino Zoroddu, Fabio Bernardini,  and  Paolo Ruggerone}
\address{INFM and Dipartimento di Fisica, Universit\`a di Cagliari,
Italy}
\author{Vincenzo Fiorentini}
\address{INFM and Dipartimento di Fisica, Universit\`a di Cagliari,
Italy\\ Walter Schottky Institut, TU M\"unchen, Garching, Germany}
\date{Submitted 24 November 2000}
\maketitle

\begin{abstract}
A number of diverse bulk properties of the zincblende and wurtzite
III-V nitrides AlN, GaN, and InN, are predicted  from first principles
within density functional theory using  the plane-wave  ultrasoft
pseudopotential method, within both the  LDA (local density) and GGA
(generalized gradient) approximations to the exchange-correlation
functional. Besides structure and cohesion, we study formation
enthalpies  (a key ingredient in predicting  defect solubilities and
surface stability), spontaneous  polarizations and piezoelectric
constants (central parameters  for nanostructure modeling), and
elastic constants. Our study bears out the relative merits of the two
density functional approaches in describing   diverse properties of
the III-V nitrides (and of the parent species N$_2$, Al, Ga, and In),
and leads us to  conclude that the GGA approximation, associated
with high-accuracy techniques such as multiprojector ultrasoft
pseudopotentials or modern all-electron methods, is to be preferred in
the study of III-V nitrides.   
\end{abstract}
\begin{multicols}{2}
\section{Introduction and method}
The III-V nitride semiconductors AlN, GaN, and InN, and their alloys are
by now well established as a strategic material system\cite{amba}
 for applications in 
high-frequency optoelectronics (LED and LASER), and high-power
electronics (e.g. HEMTs). Most of their potential in these fields is due,
respectively, to the large tunability of band gaps with alloy composition
(in principle, 1.9 to 6.2 eV), 
and to their high peak 
and saturation drift velocity, coupled with polarization-induced effects
allowing for the realization of high-density low-dimensional charge gases.
\cite{amba2}

Nitrides physics posed a number of puzzles to (and profited
considerably from) ab initio studies of  various properties and
subsystems, ranging from surfaces \cite{superfici} 
to defects,\cite{difetti} and polarization-related properties.
\cite{polarizzazione} Heralding the unusual nature of
these materials, the standard study of the structural
properties of bulk materials was a source of surprises in early
studies. For instance, since some of the earliest papers,\cite{io}
quite unusually for III-V semiconductors, the semicore 3$d$ electrons
of Ga were found to behave as valence electrons and to be essential
to describe accurately structural properties.

A major source of uncertainty, both technical and ideological in
nature,  in density-functional theory (DFT) calculations is the choice
of the exchange-correlation   functional. While LDA is used  most
commonly, the generalized gradient approximation, or GGA, has become a
close competitor in recent years. In this work, we study the effects
of using either  LDA or GGA in the prediction of the properties of
III-V nitrides.  Similar calculations have been performed only once
previously,\cite{walle} and were concerned with structural and
cohesive  properties. In this paper, we add  several new aspects to
this theme.  {\it First},  we use ultrasoft pseudopotentials, which
should in principle\cite{vand}  improve over norm-conserving 
potentials.\cite{walle}
{\it Second}, we carefully calculate formation enthalpies, which are
the cornerstone of predictions on non-stoichiometric systems relevant to
 surface reconstruction and impurity solubility.
This calculation requires us to study
 the metallic phases of Al, Ga, and In, the N$_2$
molecule, and  solid nitrogen (a molecular solid comprising
N$_2$ dimers on an hcp lattice).
{\it Third}, we evaluate the spontaneous polarization and
the piezoelectric constants of the wurtzite phase
\cite{noi-pol} in both the GGA
and LDA. We find that these quantities are 
moderately affected by the choice of
exchange-correlation, unlike most others properties.
{\it Fourth}, we evaluate a subset of
the elastic constants in LDA and GGA.

The calculations have been done using VASP (Vienna 
Ab-initio Simulation Package),\cite{vasp} which implements
the   DFT scheme within  both the LDA and GGA  approximations: we
adopted the well established  PW91 version of the GGA,\cite{PW91}
and Ceperley-Alder LDA.\cite{ca-lda} 
Ultrasoft  pseudopotentials \cite{vand}
describe the electron-ion interaction. As usual, the potentials
provided with VASP are generated for the free atom using  the
appropriate (LDA or GGA)  functional.  The pseudopotentials for 
Ga and In include respectively the semicore 3$d$ and 4$d$ states in
the valence.   A plane wave basis is used to expand the
wave-functions. We use a cut off of 350 eV, which is sufficient to
fully converge all  properties of relevance.  For k-space summation,
we use at least a Monkhorst-Pack (888) grid,
except for the N$_2$ molecule. 
Lattice constants and internal  parameters are calculated using
standard total energy calculations. Polarizations are obtained using
the Berry phase approach\cite{ksv} as in previous work.\cite{noi-pol}
Cohesive energies are calculated relative to spin-polarized
free atoms. The formation enthalpies  $\Delta H_{\rm XN}$ 
per atom pair of the XN
crystals are calculated as 
\begin{equation}
\Delta H_{\rm XN}=E_{\rm XN}-E_{\rm X}-E_{\rm N},
\end{equation}
where
$E_{\rm XN}$ is the total energy per atom pair of the compound XN,
$E_{\rm X}$ the energy per atom of bulk X = Al, Ga, and In,  and
$E_{\rm N}$ is the energy per N atom in the N$_2$ dimer or the condensed
N$_2$ phase.

\section{Parent species}
\subsection{Nitrogen: molecule and solid}
The nitrogen dimer is studied in artificial periodic conditions
in a cubic box of side 10 \AA, using the $\Gamma$ point for k-summation.
The results, listed in Table \ref{tab.N2},  agree well with other LDA
and GGA calculations. GGA shows an  overall  better agreement with
experiment. The binding energy is evaluated including the
spin-polarization  energy of the N atom (--2.89 eV), calculated  with a 
local-spin density  all-electron scalar-relativistic atomic code.\cite{bosio}

Solid nitrogen is a condensate of N$_2$ molecules.
We consider the stable phase, with vertically-oriented
 N$_2$ molecules centered at the lattice points of a close
packed hexagonal lattice. The (888)  grid is used for
k-space summation. We compare our results with experimental data from
Ref. \onlinecite{wyckoff} in Table \ref{tab.solidN2}. From the data in
Tables \ref{tab.N2} and \ref{tab.solidN2}, the binding energy per
molecule  in the condensed phase is  0.328 eV in the LDA
and 0.143 eV in the GGA. While performing well as
to the in-plane lattice constant (about $\pm$0.5\% relative
deviation), both functionals fail to some extent with the axial 
lattice parameter:
LDA underestimates it strongly ($\sim$20\%)
and GGA overestimates it ($\sim$9\%). The vertical
center-to-center intermolecular distances are 3.33 \AA\,
experimentally, 3.64 \AA\, in GGA, and 2.68 \AA\, in LDA.
This system is indeed a severe test for both functionals
because of its weak dipolar binding. GGA performs slightly
better, as expected. The binding of the N$_2$ system, already 
extremely large in reality, is overestimated appreciably
 in both approaches.

\subsection{Bulk Al, Ga, and In}
Metallic Al, Ga and In are a necessary ingredient to calculate
formation enthalpies. Al is a good free electron metal, and
 has the fcc structure. Ga is a mixed-bonding marginal metal (see e.g.
Ref. \onlinecite{bernasconi}). At ambient conditions, its
stable phase is a dimerized structure known as $\alpha$-Ga,
 a face-centered orthorombic lattice  with crystallographic
vectors ${\bf a_1}=a\hat{\bf x}$, 
${\bf a_2}=\frac{1}{2}(b\hat{\bf y}+c\hat{\bf z})$,
${\bf a_3}=\frac{1}{2}(-b\hat{\bf y}+c\hat{\bf z})$
and eight atoms per 
primitive cell, whose  positions are defined by two 
additional internal parameters $u$ and $v$.\cite{wyckoff,bernasconi}
Indium crystallizes in the monoatomic body-centered tetragonal lattice,
with lattice constants $a$ and $c$, and primitive  vectors 
${\bf a_1}=a\hat{\bf x}$,
${\bf a_2}=a\hat{\bf y}$,
${\bf a_3}=\frac{1}{2}(a\hat{\bf x}+a\hat{\bf y}+c\hat{\bf z})$.

We report our results for Al in Table \ref{tab.Al},
for Ga in Table \ref{tab.Ga}, and
for In in Table \ref{tab.In}.
In these calculations we use a cutoff of 350 eV, an (888) k-space 
mesh for Al and In, and a (12,12,12) mesh for Ga. The cohesive energy
includes the atomic  spin-polarization energy (--0.136 eV for   Al,
--0.134 eV for Ga, and --0.117 eV for In). 
As usual GGA improves somewhat the lattice constant and binding
energy. For Ga, 
both approaches are off by about the same amount in opposite
directions for $a$. In both cases,  axial ratios and internal parameters
are excellent. Our  LDA results improve somewhat over those of
Ref. \onlinecite{bernasconi}, presumably because of the explicit
treatment  of 3$d$ electrons. For In, LDA and GGA are again  off the
mark by equal and opposite  amounts for $a$. The LDA axial
ratio is slightly better than  GGA's.  In
short, the usual trend is obtained of expanded and softer lattice as
produced by GGA compared to LDA. If one is forced to choose,  GGA
generally performs better, especially in terms of cohesive energies. 
In any case, the deviations  typically are below $\pm$ 1\%,
so both approaches are quite legitimate.

\section{The nitrides}
Binary III-V nitrides occur in nature in the wurtzite
structure (the $\beta$ phase). Zinc-blende
nitrides (the $\alpha$ phase) have a slightly higher energy. It is
possible to grow epitaxially, e.g., $\alpha-$GaN on cubic substrates.
We first analyze zinc-blende (Sec. \ref{subsec.z}), then wurtzite
(Sec. \ref{subsec.w}). Our results are compared with those of
Ref. \onlinecite{walle}, where numerous other theoretical values are
provided.

\subsection{Zinc-blende AlN, GaN, InN}
\label{subsec.z}

For zinc-blende nitrides
we used the usual 350 eV cutoff and (888)  k-grid. 
 To estimate the cohesive energy,
we use the    atomic spin-polarizations indicated previously.
Our results are reported in
Tables \ref{tab.AlNzb},
\ref{tab.GaNzb}, and
\ref{tab.InNzb}, for AlN, GaN, and InN respectively.
The results confirm the by now usual behavior of GGA 
vs LDA, consisting in a softening of the lattice, which improves
lattice constant and binding energy, but worsens slightly the bulk 
modulus. Comparing the cohesive energy with that of the wurtzite phase
as discussed below, we find that zinc-blende is disfavored over
wurtzite.

\subsection{Wurtzite AlN, GaN, InN}
\label{subsec.w}

Wurtzite is a hexagonal close-packed lattice, comprising
vertically-oriented  X-N  units at the lattice sites.
The basal lattice parameter is $a$, the axial lattice parameter
is $c$. The interatomic distance in the basic unit is  described
 by an internal parameter $u$ expressed in units of the axial
ratio $c/a$.
 The ideal (i.e. for packed hard spheres) 
values of the axial ratio and internal parameter are
respectively $c/a$=$\sqrt{8/3}$ and $u=3/8$.
The crystallographic vectors of wurtzite are
{\bf a}=$a$ (1/2, $\sqrt{3}$/2, 0), 
{\bf b}=$a$ (1/2, $-\sqrt{3}$/2, 0), 
and {\bf c}=$a$ (0, 0, $c/a$). The
cartesian coordinates of the basis atoms are
(0, 0, 0), (0, 0, $u\, c$), 
$a$ (1/2, $\sqrt{3}/6$, $c/2\,a$), and
$a$ (1/2, $\sqrt{3}/6$, [$u$+1/2] $c/a$).

Our results are reported in
Tables \ref{tab.AlNwz},
\ref{tab.GaNwz}, and
\ref{tab.InNwz}, for AlN, GaN, and InN respectively.
For comparison, experimental data, and the results
of Ref. \onlinecite{walle} are also listed. 
As to structure, in all cases both the axial ratio and the internal
parameter are  non-ideal. Deviation from ideality increases from GaN
to InN to AlN. As usual GGA improves considerably lattice constant and
binding energy, at the cost of a slight overestimate of the axial
ratio. The internal parameter $u$ (alias the axial bond length) 
is well reproduced in all the various combination of materials 
and approximations.
The present GGA calculations produce lattice constants 
and internal parameters with maximum deviations from  experiment
below 0.3 \% for AlN, 0.9 \% for GaN, and 1.7 \% for InN (+1.3\% for
$a$ and +1.7\% for $c$).
In this respect, these are probably the best DFT pseudopotential 
results so far for these
materials. The improvement over previous GGA results is to be
attributed to the use of ultrasoft, multiprojector
pseudopotentials.\cite{vand} By the same token, it is   
quite likely that all-electron calculations using the same GGA
parameterization may improve the agreement further, especially for
InN.

Calculated cohesive energies overestimate, as usual, the experimental value.
GGA  corrects in part the LDA overbinding, and exhibits
 better agreement. Comparing the cohesive energies of the 
zinc-blende and wurtzite phases, as already mentioned, 
we find  wurtzite to be  energetically favored
over zincblende.  The predicted difference per atom pair
 between the  two phases is 189 meV (LDA) and 164 meV (GGA) for AlN, 
 17 meV (LDA) and 16 meV (GGA) for GaN,
17 meV (LDA) and 15 meV (GGA) for InN.

 Good results are also achieved for the formation enthalpies.
These were obtained using  the energy per N atom in the solid-N$_2$
phase: if the N$_2$ molecule is assumed as reference instead,
one half of the binding energy of solid-N$_2$ (i.e. 0.164 eV in LDA,
 and 0.071 eV in GGA) must be added to the values in the Tables.
The reported values are referred to an atom pair.
The available formation enthalpy  measurements for the nitrides
 are not very recent,\cite{Ans,properties} and their spread
 is considerable (this is presumably also the case for cohesive
energies, which are measured as heats of vaporization).  The calculated
GGA formation enthalpies are in general agreement with experiment for AlN 
and GaN. For InN,  GGA overcorrects
the LDA overbinding and  gives a positive value. This problem is due
to InN itself, and not to In or N parent phases: indeed, quite
unusually,  the calculated cohesive energy {\it underestimates} the
experimental value. We are not aware of other formation enthalpy
calculations for InN. The  issue is open to further
investigation, especially by all-electron methods. 

In Table \ref{tab.new}, we report for each of the nitrides
the  spontaneous polarization in the equilibrium structure, 
the dynamical effective charges, the
 piezoelectric constants, and a subset of
elastic constants relevant to symmetry-conserving strains. The reason
for collecting these data in one Table is that they provide an almost
self-contained set of input data for the  simulation of nanostructures
made of wurtzite nitrides. 
The only additional data needed are the static dielectric constants,
which were reported elsewhere.\cite{noi-prl}
In the last column we report the proper piezoelectric constant
$e_{31}^{\rm p}$.
As discussed recently,\cite{van2,noi-anti} this value should be compared 
with experiments involving current flow across the sample, whereas
the ``improper'' constant $e_{31}$ is relevant to systems in
depolarizing fields such as nitride
nanostructures.\cite{polarizzazione}

It is not infrequent to hear the incorrect statement that the
 spontaneous polarization is non vanishing in wurtzite because of 
 structural non-ideality. In actual fact, a  non-vanishing 
polarization is allowed on symmetry grounds \cite{burns} in
the {\it ideal} wurtzite structure as well. Indeed, we find 
that the calculated Berry-phase polarization 
in the ideal structure is indeed non-zero:     --0.032 C/m$^2$ in AlN,
    --0.018 C/m$^2$ in GaN, and    --0.017
C/m$^2$ in InN. These values are smaller (by a factor of 2 to 3)
than the actual ones for non-ideal structures (Table
\ref{tab.new}). This confirms the intuitive idea that non-ideality, and
especially changes in $u$, can increase polarization substantially,
and indicates that an accurate determination of the structure is
mandatory to obtain reliable polarization values.\cite{noi-anti}

Theoretical predictions on  polarization properties were shown
to compare quite favorably with experimental evidences in 
various papers (see e.g. Refs. \onlinecite{amba2},
\onlinecite{polarizzazione}, and
\onlinecite{walter}). It should be noted, however, that
 the link between polarization and the observed quantities, typically
optical shifts or densities of  mobile charge, is rather indirect and
affected by uncertainties due to issues of nanostructure 
design, material quality, and reverse modeling. 
Thus, comparison with experiment  does not yet allow 
a clear-cut evaluation of the performance of LDA vs GGA.
The  recently reported\cite{noi-nl} non-linear behavior of the
polarization in nitride alloys is an additional source of
uncertainty. 

The LDA elastic constants are in fair agreement with those of
Wright.\cite{wright} The GGA constants are smaller,
as is to be expected given the general tendency of GGA to produce a
softer lattice. According to elasticity theory, the axial strain
induced in wurtzite by an in-plane (e.g. epitaxial)   strain
$\epsilon_1$ is $\epsilon_3 = -2\,\epsilon_1\, C_{31}/C_{33} = R\,
\epsilon_1$. The quantity $R$ is thus relevant to epitaxial nitride
systems, and it is  reported in Table \ref{tab.new}. Several
experimental data for the elastic constants and $R$  are compiled in 
Ref. \onlinecite{wright}. The considerable spread of those data
 does not allow a definite conclusion about whether GGA produces a
systematically improved agreement with experiment over LDA in this
respect.

\section{Summary}
In conclusion, GGA and LDA calculations for III-V nitrides suggest
an overall improvement of the predicted properties in the former
approximation. In particular, the structural parameters are
extremely accurate, with deviations from experiment
below 0.3 \% for AlN, 0.9 \% for GaN, and 1.7 \% for InN.
Cohesive energies and formation enthalpies are in fair to excellent
agreement with experiment; the only clear-cut failure (in the GGA
approximation) is the positive formation enthalpy of InN.
Elastic properties follow the expected trends of GGA vs LDA behavior;
due to uncertainties in the experimental data, comparison with
experiment does not provide definite support to one or the other
approximation. 
Polarization properties are moderately sensitive to the
exchange-correlation functional, as long as  the latter predicts the
correct structure (especially, the correct internal parameter
$u$). For these properties too,  comparison with experiment is 
indirect and affected  by many sources of uncertainty, and does not 
support one or the other approach.
 Concerning the cohesion and structure of parent species
(N$_2$, Ga, Al, In), only in the case of condensed N$_2$ do we find 
major discrepancies with experiment.  
In the light of the present results, 
our conclusion is that the GGA approximation should be preferred 
in density-functional studies of III-V nitrides. Of course, the choice of
one or the other approximation may depend on the specific problem
being addressed.

\subsection*{Acknowledgements}

We thank Dr. A. Bosin for atomic calculations.
Work at Cagliari University is supported in part by
MURST--Cofin99, and by INFM Parallel Computing Initiative.
VF thanks the Alexander von Humboldt-Stiftung for support
of his stay at the Walter Schottky Institut, and O. Ambacher for 
pointing out recent and accurate structural data.


\narrowtext\begin{table}[h]
\caption{Bond length, vibrational frequency, and binding
 energy of the N$_2$ dimer.}
\label{tab.N2}
\begin{tabular}{llll}
   &  $d$ (\AA) & $\omega$ (THz) & E$_b$ (eV) \\
\hline
LDA    &  1.107  & 464.3      & --11.332\\
GGA    &  1.113  & 442.8      & --10.558\\
Experiment\tablenote{Ref.\onlinecite{walle}}&  1.10   & 444.8      &  --9.9  \\
\end{tabular} 
\end{table}
\begin{table}[h]
\caption{Structural parameters and binding energy
 per molecule of hexagonal solid N$_2$}
\label{tab.solidN2}
\begin{tabular}{llll}
  & $a$ (\AA)  &   $c/a$    &    E$_b$ (eV)   \\    
\hline
LDA (present)         &  4.0205    &   1.3311          &   --11.660  \\
GGA (present)       &  4.0633    &   1.7929          &   --10.701  \\
Experiment\tablenote{Ref. \onlinecite{wyckoff}}   &  4.039     &   1.6514          \\
\end{tabular}

\end{table}
             

\begin{table}
\caption{Lattice constant, binding energy, and bulk
 modulus of bulk fcc Al.}
\label{tab.Al}
\begin{tabular}{llll}
  &   $a$ (\AA)   &  E$_b$ (eV)  &  B(MBar)  \\
\hline
LDA (present)        & 3.9809   & --4.064  &  0.766 \\
GGA (present)        & 4.0491   & --3.561  & 0.689   \\
Experiment\tablenote{Ref. \onlinecite{Kittel}}
 &  4.05      & --3.39   &  0.773\\
\end{tabular}
\end{table}
\begin{table}
\caption{Lattice constant, binding energy, axial ratios,
and internal parameter (units of $c$) of $\alpha$-Ga.}
\label{tab.Ga}
\begin{tabular}{lllllll}
  &  $a$ (\AA)  & $b/a$ & $c/a$ & $u$ 
& $v$ &  E$_b$ (eV)  \\
\hline
LDA (present)        &  4.4365    &    0.9985   &    1.6856     &   0.0816
&  0.1577          & --3.484 \\
LDA\tablenote{Ref.\onlinecite{bernasconi}} &  4.377    &    0.994  
 &    1.688     &   0.0803
&  0.1567          &  \\
GGA (present)       &   4.5962    &    0.9917     &    1.6961     &   0.0834 
& 0.1559        & --2.796  \\
Experiment$^{\rm a}$    &   4.51     &    1.0013      &    1.695 
     &    0.0785        &  0.1525        &  \\
\end{tabular}
\end{table}

\begin{table}
\caption{Lattice constant, axial ratio, and binding energy of bulk In.}
\label{tab.In}
\begin{tabular}{llll}
    &   $a$ (\AA)   &     $c/a$     &  E$_b$ (eV)   \\
\hline
LDA (present)   &  3.1861  &  1.5348  &  --3.116  \\
GGA (present)   &  3.2958  &  1.5448  &  --2.470  \\
Experiment        &  3.244\tablenote{Ref. \onlinecite{wyckoff}}   &
1.5222$^{\rm a}$ & \\
\end{tabular}
\end{table}

\begin{table}
\caption{Lattice constant, binding energy, 
 and formation enthalpy $\Delta H$
 of zinc-blende  AlN.}
\label{tab.AlNzb}
\begin{tabular}{lllc}
&  $a$ (\AA)    &  E$_b$ (eV) & $\Delta H$ (eV) \\
LDA (present)       &  4.332   & --13.347&  --3.449 \\
LDA\tablenote{Ref. \onlinecite{walle}}       &  4.310   & --13.242&   \\
GGA (present)       &  4.390   & --11.907&  --2.975 \\
GGA$^{\rm a}$       &  4.394   & --11.361&   \\
Experiment\tablenote{Ref. \onlinecite{properties}}      &  4.37    &  \\ 
\end{tabular}
\end{table}
\begin{table}   
\caption{Lattice constant, binding energy, 
 and formation enthalpy $\Delta H$
 of zinc-blende GaN.}
\label{tab.GaNzb}
\begin{tabular}{llrc}
    &  $a$ (\AA)  &  E$_b$ (eV)       & $\Delta H$ (eV)\\
\hline
LDA (present)          & 4.446  & --10.982&     --1.689\\
LDA\tablenote{Ref.\onlinecite{walle}} & 4.518  & --10.179     &       \\
LDA\tablenote{Ref.\onlinecite{io}} & 4.466  & --10.880     &       \\
GGA (present)          & 4.538  & --9.249   & --1.102\\
GGA$^{\rm a}$ & 4.590  & --8.253     &       \\ 
Experiment\tablenote{Ref. \onlinecite{properties}}             & 4.519  &
 &   \\   
\end{tabular}
\end{table}
\begin{table}
\caption{Lattice constant, binding energy, 
 and formation enthalpy $\Delta H$ 
 of zinc-blende InN.}
\label{tab.InNzb}
\begin{tabular}{lllc}
 &  $a$ (\AA) &  E$_b$ (eV)    &$\Delta  H$ (eV)  \\
\hline
LDA (present)         & 4.964  & --9.232  &   --0.282 \\
LDA\tablenote{Ref.\onlinecite{walle}}  & 5.004  & --8.676    &        \\
GGA (present)        & 5.067  & --7.680    & 0.140 \\
GGA$^{\rm a}$  & 5.109  & --6.855 &        \\
Experiment\tablenote{Ref. \onlinecite{properties}}       & 4.98 
&   \\
\end{tabular}
\end{table}
\begin{table}
\caption{Lattice constant, axial ratio, internal parameter, and
formation enthalpy 
of wurtzite AlN.}
\label{tab.AlNwz}
\begin{tabular}{lllllc}
   &   $a$ (\AA)     &  $c/a$     &    $u$      &
 E$_b$ (eV)& $\Delta H$ (eV)      \\
\hline
LDA (present)          &   3.0798   &  1.5995   &  0.3821   &--13.536&
--3.642 \\
LDA\tablenote{Ref.\onlinecite{walle}}  &   3.057   &  1.617   &
0.3802   &--13.286& \\ 
GGA (present) &   3.1095   &  1.6060   &  0.3819   &--12.071&
--3.142 \\
GGA$^{\rm a}$   &   3.113   &  1.6193   &  0.3798   &--11.403&  \\
Experiment\tablenote{Ref.\onlinecite{Tanaka}}         &   3.1106 &
1.6008 & 0.3821\tablenote{Ref.\onlinecite{Les}}  & --11.669$^{\rm a}$&
 --3.13\tablenote{Ref.\onlinecite{Ans}}
--2.53\tablenote{Ref.\onlinecite{properties}}\\
\end{tabular}
\end{table}
\begin{table}
\caption{Lattice constant, axial ratio, internal parameter, and formation enthalpy
of wurtzite GaN.}
\label{tab.GaNwz}
\begin{tabular}{llllrc}
   &   $a$ (\AA)   &    $c/a$    &    $u$     & E(ev) &
 $\Delta H$ (eV)  \\ 
\hline
LDA (present)          &   3.1461 &  1.6273  &  0.3768  &--10.999& --1.685 \\
LDA\tablenote{Ref.\onlinecite{walle}}  &   3.193  &  1.634   &  0.376   & --10.187\\
GGA (present)          &   3.1986 &  1.6339  &  0.3772  & --9.265 & --1.118 \\
GGA$^{\rm a}$  &   3.245  &  1.632   &  0.3762  & --8.265 \\
Experiment\tablenote{Ref.\onlinecite{Les}}         &   3.1890 &
1.6263 &  0.377 & --9.058$^{\rm a}$ &
--1.08\tablenote{Ref.\onlinecite{Ans}}
--1.91\tablenote{Ref.\onlinecite{properties}}\\  
\end{tabular}
\end{table}
\begin{table}
\caption{Lattice constant, axial ratio, internal parameter, and
formation enthalpy 
of wurtzite InN.}
\label{tab.InNwz}
\begin{tabular}{lllllc}
   &   $a$ (\AA) &    $c/a$        &    $u$    & E$_b$ (eV) &
$\Delta H$ (eV)   \\
\hline
LDA (present)          &   3.5218 &  1.6121  &  0.3791  & --9.249& --0.303 \\ 
LDA\tablenote{Ref.\onlinecite{walle}}  &   3.544  &  1.626   &  0.377
& --8.694\\ 
GGA (present)          &   3.5848 &  1.6180  & 0.37929  & --7.695& 0.125 \\
GGA$^{\rm a}$  &   3.614  &  1.628   & 0.377    & --6.872\\
Experiment\tablenote{Ref.\onlinecite{pasz}}&
3.538  &  1.6119&      
    & --7.970\tablenote{Ref.\onlinecite{properties}}& 
--0.21\tablenote{Ref.\onlinecite{Ans}}
--1.36\tablenote{Ref.\onlinecite{new}}\\  
\end{tabular}
\end{table}

\begin{table}
\caption{Spontaneous polarization (C/m$^2$),
piezoelectric constants (C/m$^2$), 
 dynamical charges, elastic constants (GPa),
and the ratio $R$=$-2\, C_{31}/C_{33}$  (see text)
 of wurtzite nitrides, as obtained in
the LDA and GGA approximation.
The last column reports the proper $e_{31}$ piezoelectric constant.}
\label{tab.new}
\begin{tabular}{lcccccccc}
\multicolumn{1}{c}{}&
\multicolumn{1}{c}{P$_{\rm}$}&
\multicolumn{1}{c}{Z$^{*}$} &
\multicolumn{1}{c}{$e_{33}$}&
\multicolumn{1}{c}{$e_{31}$}&
\multicolumn{1}{r}{$C_{33}$}&
\multicolumn{1}{c}{$C_{31}$}&
\multicolumn{1}{c}{$R$}&
\multicolumn{1}{c}{$e_{31}^{\rm p}$}
\\
\hline
AlN &  & &  &  &  & \\
LDA  &  --0.100 &     2.652 &   1.80 &    --0.64  &   384  &   111& --0.578 & 
--0.74\\
LDA\tablenote{Ref.\onlinecite{wright}}   &   &       &    &    & 373    &  
108 & --0.579 & \\
GGA  &  --0.090 &     2.653 &   1.50 &    --0.53  &   377  &   94 &
--0.499 & --0.62\\
\hline
GaN &  & &  &  &  & \\ 
LDA   & --0.032   &   2.51    &  0.85  &  --0.44   &  415  &   83 &
--0.400 & --0.47\\
LDA $^{\rm a}$   &   &       &    &    & 405    & 103  & --0.508
& \\
GGA   & --0.034   &   2.67    &  0.66  &  --0.34   &  354  &   68 & --0.384
&  --0.37\\
\hline
InN &  & &  &  &  & \\
LDA   & --0.041  &    3.045   &  1.09  &  --0.52  &   233  &  88 & 
--0.755 &
--0.56\\
LDA $^{\rm a}$   &   &       &    &  &224  & 92    &  --0.821 & \\
GGA   & --0.042  &    3.105   &  0.81  &  --0.41  &   205  &   70 &
--0.683 &--0.45\\
\end{tabular}
\end{table}
\end{multicols}
\end{document}